\documentclass[aps,preprint,showpacs]{revtex4}
\usepackage{epsfig}
\begin{document}
\title{Market memory and fat tail consequences in option pricing on the expOU stochastic volatility model}
\author{Josep Perell\'o~\cite{email}}
\affiliation{Departament de F\'{\i}sica Fonamental, Universitat de Barcelona,\\ Diagonal, 647, E-08028 Barcelona, Spain}
\date{\today}

\begin{abstract}
The expOU stochastic volatility model is capable of reproducing fairly well most important statistical properties of financial markets daily data. Among them, the presence of multiple time scales in the volatility autocorrelation is perhaps the most relevant which makes appear fat tails in the return distributions. This paper wants to go further on with the expOU model we have studied in Ref.~\cite{masoliver} by exploring an aspect of practical interest. Having as a benchmark the parameters estimated from the Dow Jones daily data, we want to compute the price for the European option. This is actually done by Monte Carlo, running a large number of simulations. Our main interest is to ``see" the effects of a long-range market memory from our expOU model in its subsequent European call option. We pay attention to the effects of the existence of a broad range of time scales in the volatility. We find that a richer set of time scales brings to a higher price of the option. This appears in clear contrast to the presence of memory in the price itself which makes the price of the option cheaper.

Keywords: stochastic volatility, option pricing, long memory

\end{abstract}
\pacs{89.65.Gh, 02.50.Ey, 05.40.Jc, 05.45.Tp}

\maketitle

\section{Introduction}

The model, suggested by Bachelier in 1900 as an ordinary random walk and redefined in its final version by Osborne in 1959~\cite{osborne}, presupposes a constant ``volatility'' $\sigma$, that is to say, a constant diffusion coefficient $D=\sigma^2$. However, and especially after the 1987 crash, there seems to be empirical evidence, embodied in the so-called ``stylized facts'', that the assumption of constant volatility does not properly account for important features of markets~\cite{cont,plerou,bouchaud}. It is not a deterministic function of time either (as might be inferred by the evidence of non stationarity in financial time series) but a {\it random variable}. In its more general form one therefore may assume that the volatility $\sigma$ is a given function of a random process $Y(t)$, {\it i.e.}, $\sigma(t)=\phi(Y(t))$. 

At late eighties different stochastic volatity (SV) models [Brownian motion with random diffusion coefficient] were presented for giving a better price to the options somewhat ignoring their ability to reproduce the real price time series~\cite{fouquebook}. More recently SV models have been suggested by some physicists as good candidates to account for the so-called stylized facts of speculative markets~\cite{dragulescu,perello,perello1,perello2,silva,duarte}. In the context of mathematical finance, we mainly have three models: the Ornstein-Uhlenbeck (OU)~\cite{perello2}, the Heston~\cite{dragulescu} and the exponential Ornstein-Uhlenbeck model~\cite{fouque}. We have recently tried to decide which model works better~\cite{perello,perello1}. Very recently we have studied the exponential Ornstein-Uhlenbeck stochastic volatility model~\cite{fouque} and observed that the model shows a multiscale behavior in the volatility autocorrelation~\cite{masoliver}. It also exhibits a leverage correlation and a probability profile for the stationary volatility which are consistent with market observations. All these features seem to make the model more complete than other stochastic volatility models also based on a two-dimensional diffusion. It is worth to mention that there has been some more sophisiticated models based on a three-dimensional diffusion process that reproduce the intrincate set of memories in market dynamics~\cite{vicente,perello3}. Indeed, coming from multifractality framework, there are recent nice papers~\cite{bacry} with promising features to provide even a better description that the one by the common stochastic volatility models~\cite{sornette}.

\section{The volatility model}

Let us briefly summarize the main definitions and the properties of the exponential Ornstein-Uhlenbeck stochastic volatility model~\cite{masoliver}. The model consists in a two-dimensional diffusion process given by the following pair of It\^o stochastic differential equations (SDEs):
\begin{eqnarray}
\dot{X}(t)&=&me^{Y(t)}\xi_1(t)\label{1d}\\
\dot{Y}(t)&=&-\alpha Y(t)+k\xi_2(t),
\label{2d}
\end{eqnarray}
where dots represent time derivative. The variable $X(t)$ is the undrifted log-price or zero-mean return defined as $\dot{X}=\dot{S}(t)/S(t)-\left\langle\dot{S}(t)/S(t)\right\rangle$, where $S(t)$ is a the asset price. The parameters $\alpha$, $m$, and $k$ appearing in Eq.~(\ref{2d}) are positive and nonrandom quantities. The two noise sources of the process are correlated Wiener processes, {\it i.e.}, $\xi_i(t)$ are zero-mean Gaussian white noise processes with cross correlations given by
\begin{equation}
\left\langle\xi_i(t)\xi_j(t')\right\rangle=\rho_{ij}\delta(t-t'),
\label{rho}
\end{equation}
where $\rho_{ii}=1$, $\rho_{ij}=\rho$ $(i\neq j, -1\leq\rho\leq 1)$. In terms of
the proces $Y(t)$ the volatility is given by
\begin{equation}
\sigma(t)=me^{Y(t)}.
\label{sigma}
\end{equation}
It is worth to mention that multifractals models also considers a random variable which describes the logarithm of the volatility~\cite{bacry,sornette}.

\begin{figure}[t]
\centerline{\epsfig{file=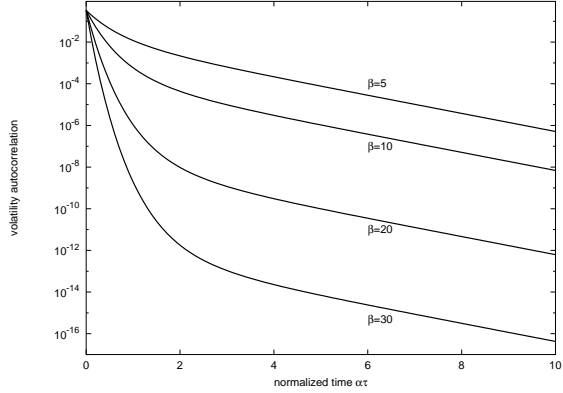,width=7.6cm}}
\caption{Volatility autocorrelation~(\ref{volcorfin}) for different values of $\beta$. The characteristic time scale for the long time corresponds to $1/\alpha$ while short time scale in leverage is related to $1/k^2$. For a large value of $\beta$ we observe a richer multiple time scale behaviour.}
\label{fig2}
\end{figure}

Among the most important results of the expOU, we must stress the following three~\cite{masoliver}. First of all the stationary volatility pdf which is a log-normal distribution being quite consistent with empirical data~\cite{bouchaud} 
\begin{equation}
p_{st}(\sigma)=\frac{1}{\sigma\sqrt{2\pi\beta}}\exp\left\{-\ln^2(\sigma/m)/2
\beta\right\},
\label{statpdf}
\end{equation}
where $\beta=k^2/2\alpha$. Notice that the stationary distribution broadens the tails as we increase the value of $\beta$. Secondly, we have the leverage effect~\cite{bouchaud}
\begin{equation}
{\cal L}(\tau)=\frac{\left\langle dX(t)dX(t+\tau)^2\right\rangle}{\left\langle
dX(t)^2\right\rangle^2}=(2\rho k/m)\exp\left\{-\alpha\tau+2\beta(e^{-\alpha\tau}-
3/4)\right\}H(\tau),
\label{leveragefin}
\end{equation}
where $H(\tau)$ is the Heaviside step function. One can show that
\begin{equation}
{\cal L}(\tau)\simeq \frac{2}{m}\rho ke^{\beta/2} e^{-k^2\tau}\qquad(
\alpha\tau\ll 1),
\label{levershort}
\end{equation}
while leverage is negligible for large times. These approximations hold only if we take $\beta>1$.
Thirdly, and perhaps the most important, the volatility autocorrelation
\begin{equation}
{\rm Corr}(\tau)=\frac{\langle dX(t)^2dX(t+\tau)^2\rangle-\langle dX(t)^2\rangle^2}{
\langle dX(t)^4\rangle-\langle dX(t)^2\rangle^2}=\frac{\exp[4\beta e^{-\alpha\tau}]-1}{3e^{4\beta}-1}.
\label{volcorfin}
\end{equation}
which expanded in the right way allow us to observe a cascade of exponentials
\begin{equation}
{\rm Corr}(\tau)=\frac{1}{3e^{4\beta}-1}\sum_{n=1}^{\infty}\frac{(4\beta)^n}{n!}	e^{-n
\alpha\tau}.	\label{volcorexpand}
\end{equation}
This expression indicates that there are infinite time scales between the limiting behaviours
\begin{equation}
{\rm Corr}(\tau)\approx\frac{4\beta}{3e^{4\beta}-1}e^{-\alpha\tau} \quad (\alpha\tau\gg 1), \qquad
{\rm Corr}(\tau)=\frac{1}{3e^{4\beta}-1}\left[e^{4\beta-2k^2\tau}-1 \right]+\mbox{O}\left
(\alpha^2\tau^2\right)
\end{equation}
As one can observe the characteristic time scale for the long time might be associated to $1/\alpha$ while short time scale in leverage is related to $1/k^2$ (see Fig.~\ref{fig2}). The distance between the smallest time scale and the largest is given by $\beta=k^2/2\alpha >1$. The bigger $\beta$, the larger is the distance and the richer is the cascade of multiple time scales of the process. Even more, as we enlarge the distance between smaller time scale and larger time scale, we also get a broader lognormal distribution (cf. Eq.~(\ref{statpdf})) for the volatility and a fatter distribution fo the returns.

\section{Option pricing}

Having in mind latter expressions and main messages concerning $\beta$, we can start looking at the inferrences of the volatility long range memory in the option price. An European option is a financial instrument giving to its owner the right but not the obligation to buy (European call) or to sell
(European put) a share at a fixed future date, the maturity time $T$, and
at a certain price called exercise or striking price $K$. In fact, this
is the most simple of a large variety of derivatives contracts. In a certain sense, options are a security for
the investor thus avoiding the unpredictable
consequences of operating with risky speculative stocks. 

The payoff of the European call contract at maturity date $T$ is
$$
S(T)-K \quad {\rm if }\quad S(T)>K,\qquad  {\rm and} \quad 0 \quad {\rm otherwise}.
$$ 
To compute the price of the call we can use the average (on an equivalent martingale measure) over 
a large set of simulated paths which can be written mathematically as follows
\begin{equation} 
C_T(S)=E^*\left[\left. e^{-rT}(S(T)-K)^+\right|S(t=0)=S\right].
\end{equation}
To do so we assume the process defined in the pair of Eqs.~(\ref{1d})-(\ref{2d}) but with a drift equal to the risk free interest ratio $r$ and with a market price of risk for the volatility component set to be 0. That is: keeping the current measure for the dynamics of the volatility. There are many subtleties behind the evaluation of the volatility market price of risk and we address the reader to the book by Fouque et al.~\cite{fouquebook} to get a deeper knowledge on this issue.

\begin{figure}[t]
\centerline{\epsfig{file=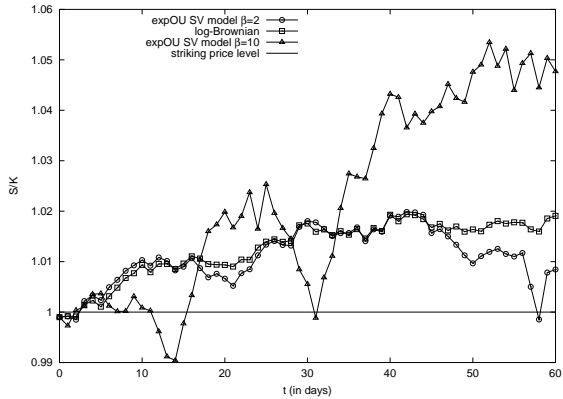,width=7.6cm}}
\caption{Random path of the log-Brownian motion compared to a couple of expOU SV model paths with different $\beta$'s. Fluctuations increases wildly with $\beta$. Price time series take returns with a non-zero drift $r=1.845 \mbox{ day}^{-1}$ (5\% annual risk free interest rate ratio). Rest of the parameters are $k=0.18 \mbox{ day}^{-1/2}$, $m=0.00141 \mbox{ day}^{-1/2}$ and for the log-Brownian motion (constant volatility) we take $\sigma=m$. All series take an initial stock price $S=0.999K$. }
\label{fig1}
\end{figure}

We thus simulate the random path assuming one day timesteps. Figure~\ref{fig1} shows the different paths observed for the SV model and the log-Brownian one using $m$ as a volatility o diffusion coefficient (cf. Eq.(~\ref{sigma})). Repeating the paths 1,000,000 times over the same time window, we can get an approximate idea of the correction to the classic Black-Scholes price formula. We recall that Black-Scholes assumes a log-Brownian motion for the underlying whose price is well-known and has the analytical solution:
\begin{equation}
C_{\rm BS}(S,t)=SN(d_1)-Ke^{-r(T-t)}N(d_2) \qquad (0\leq t\leq T),
\label{formula}
\end{equation}
where $N(z)$ is the probability integral 
\begin{equation}
N(z)=\frac{1}{\sqrt{2\pi}}\int_{-\infty}^z e^{-x^2/2}dx,
\label{dbs}
\end{equation}
and its arguments are 
\begin{equation}
d_1=\frac{\ln(S/K)+(r+\sigma^2/2)(T-t)}{\sigma\sqrt{T-t}}, \qquad
d_2=d_1-\sigma\sqrt{T-t}.
\label{dbs1}
\end{equation}

In contrast with some contributions from mathematical finance, we are not inserting the parameters values blindly nor providing a large collection of parameters where it is quite hard to intuit the meaning and the effects of each parameters. We take the parameters set in Ref.~\cite{masoliver} for the Dow Jones daily index as a benchmark. The parameters derived gives an opposite approach to the one already performed by Fouque etal.~\cite{fouque} for the expOU. They are focused in analytical asymptotic results with the cases where $\beta<1$ but the problem is that with this restriction one does not have the desired cascade of time scales in the volatility autocorrelation.

\begin{figure}[t]
\centerline{\epsfig{file=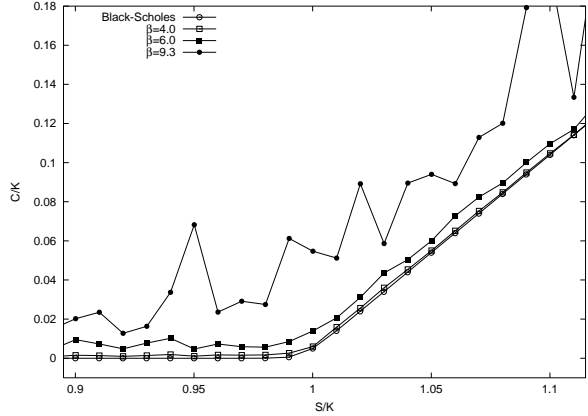,width=8cm}}
\caption{The European call option in terms of the moneyness $S/K$. We show the Black-Scholes formula and the expOU SV model for several values of the parameters. The computation with the expOU is performed by running 1,000,000 paths and average over their final premium. Results become very noisy for $\beta$ around 10.}
\label{fig3}
\end{figure}

We recall that $m$ is the volatility level assuming no auxilary diffusion process for the volatility, $\rho$ gives the asymetry between positive and negative return change and the appropriate sign for the leverage effect. And finally, short range memory time scale is measured by $k^2$ which appears to be of the order of few weeks (10-20 days). We will now focus on the effect of the largest time scale memory $\alpha^{-1}$ which is considered to be of the order of years. We have shown in Ref.~\cite{masoliver} that a good approximate solution for reproduce the memories of the market is performed taking $k^2=1.4\times 10^{-2} \mbox{ day}^{-1}$, $\rho=0.4$ and $m=1.5 \times 10^{-5} \mbox{ day}^{-1/2}$. The comparison between shortest and largest time scale is provided with $\beta$ and at least for the Dow Jones daily data this is around 3.8. 

\begin{figure}[t]
\centerline{\epsfig{file=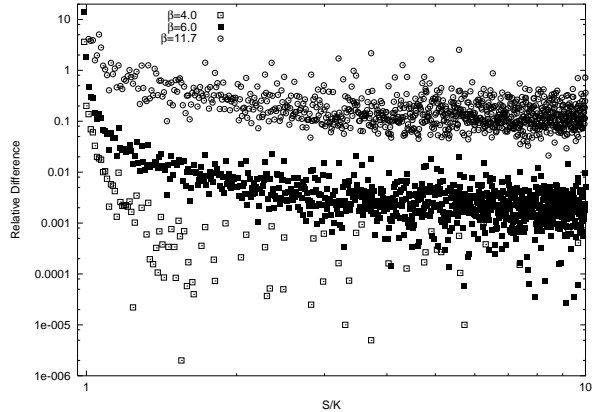,width=8cm}}
\caption{The relative difference between the expOU call option and the European call option in terms of the moneyness $S/K$. We show the difference for several values of the parameters. The computation with the expOU is performed by running 1,000,000 paths and average over their final premium.}
\label{fig4}
\end{figure}

In Figure~\ref{fig3} we plot the call price for several values of $\beta$ averaging over the initial volatility with the lognormal distribution~(\ref{statpdf}). We take maturity date at 20 days and we represent the option assuming that the risk free interest ratio to 5\% per year. Even for such a small time horizon (much smaller than the volatility memory), we get important price differences. And in any case the longer the memory is the more expensive is the option. This can also be quantified by the relative price between new option price and the Black-Scholes one, that is:
\begin{equation}
\mbox{Relative difference}=\frac{C(S,t)}{C_{\rm BS}(S,t)}-1
\end{equation}
The results are plotted in Fig.~\ref{fig4}. One observe that the price difference becomes more important with a monotonic increase in terms of $\beta$. These differences may become relatively very important for small moneyness $S/K$. And the decay of the relative difference for larger distances with respect to the striking price $K$ is becoming slowler with a higher value of $\beta$. We have tested the results with different maturity dates and with different values for $k,\rho$, and $m$ with similar conclusions.

In a previous paper~\cite{option}, we also have been studying the effects of the memory in the option price. In that case, however, we had memory in the price itself breaking the efficient market hypothesos. We had observed that the call became cheaper with this kind of memory even if this of only one day. The presence of memory in the volatility has opposite consequences. This paper has aimed to insist in the fact that the memory persistence in volatility affects the price making this to be higher.

\begin{acknowledgments}
I wish to warmly thank Jaume Masoliver for useful discussions on the expOU modeling. 
This work has been supported in part by Direcci\'on General de Investigaci\'on
under contract No. BFM2003-04574 and by Generalitat de Catalunya under contract
No. 2000 SGR-00023.
\end{acknowledgments}

\end{document}